\documentclass[preprint]{aastex}
\usepackage{psfig}
\begin{document}
\slugcomment{To Appear in ApJ}
\title{The Survey of Nearby Nuclei with STIS (SUNNS): Emission-Line Nuclei
at {\em Hubble Space Telescope} Resolution\altaffilmark{1}}

\author{Joseph C. Shields\altaffilmark{2}, Hans-Walter Rix\altaffilmark{3},
Marc Sarzi\altaffilmark{4}, Aaron J. Barth\altaffilmark{5}, 
Alexei V. Filippenko\altaffilmark{6}, Luis C. Ho\altaffilmark{7}, 
Daniel H. McIntosh\altaffilmark{8}, Gregory Rudnick\altaffilmark{9, 10}, and 
Wallace L. W. Sargent\altaffilmark{11}}

\altaffiltext{1}{Based on observations with the NASA/ESA {\sl Hubble Space
Telescope} obtained at the Space Telescope Science Institute, which is
operated by the Association of Universities for Research in Astronomy,
Inc., under NASA contract NAS5-26555.}
\altaffiltext{2}{Physics \& Astronomy Department, Ohio University, Athens,
OH 45701}
\altaffiltext{3}{Max-Planck-Institut f\"ur Astronomie, K\"onigstuhl 17,
Heidelberg D-69117, Germany}
\altaffiltext{4}{Centre for Astrophysics Research, University of
Hertfordshire, AL10 9AB Hatfield, United Kingdom}
\altaffiltext{5}{Physics \& Astronomy Department, University of California,
4129 Frederick Reines Hall, Irvine, CA 92697-4575}
\altaffiltext{6}{Astronomy Department, University of California, Berkeley,
CA 94720-3411}
\altaffiltext{7}{The Observatories of the Carnegie Institution of Washingon,
813 Santa Barbara St., Pasadena, CA 91101}
\altaffiltext{8}{Astronomy Department, University of Massachusetts, Amherst,
MA 01003}
\altaffiltext{9}{National Optical Astronomy Observatory, P.O. Box 26732,
Tucson, AZ 85726-6732}
\altaffiltext{10}{Goldberg Fellow}
\altaffiltext{11}{Astronomy Department, Caltech 105-24, Pasadena, CA 91125}

\begin{abstract}

We present results from a program of optical spectroscopy for 23
nearby galaxies with emission-line nuclei.  This investigation takes 
advantage of the spatial resolution of the {\em Hubble Space Telescope} 
to study the structure and energetics of the central $\sim 10 - 20$ pc,
and the resulting data have value
for quantifying central black hole masses, star formation histories,
and nebular properties.  This paper provides a description of the
experimental design, and new findings from the study of emission
lines.  The sample targets span a range of nebular spectroscopic
class, from \ion{H}{2} to Seyfert nuclei.  This data set and the
resulting measurements are unique in terms of the sample size, the
range of nebular class, and the investigation of physical scales
extending down to parsecs.  The line ratios indicative of nebular
ionization show only modest variations over order-of-magnitude
differences in radius, and demonstrate in a systematic way that
geometrical dilution of the radiation field from a central source
cannot be assumed as a primary driver of ionization structure.
Comparisons between large- and small-aperture measurements for the
\ion{H}{2}/LINER transition objects provide a new test that challenges
conventional wisdom concerning the composite nature of these systems.
We also list a number of other quantitative results that are of
interest for understanding galaxy nuclei, including (1) the spatial
distribution/degree of concentration of H$\alpha$ emission as a
function of nebular type; (2) the radial variation in electron density
as a function of nebular type; and (3) quantitative broad H$\alpha$
estimates obtained at a second epoch for these low-luminosity nuclei.
The resulting measurements provide a new basis for comparing the
nuclei of other galaxies with that of the Milky Way.  We find that the
Galactic Center is representative across a wide span of properties as
a low-luminosity emission-line nucleus.

\end{abstract}


\section{Introduction}

The physical properties of galaxy nuclei are of larger interest in
light of growing indications that evolution in the central few parsecs
is linked to the global properties of the surrounding galaxy.  This
connection is evident particularly in correlations between the mass of
a central black hole, and bulge mass (e.g., Kormendy \& Richstone 1995;
H\"aring \& Rix 2004) or velocity dispersion (Gebhardt et al. 2000;
Ferrarese \& Merritt 2000).  Star formation in the centers of galaxies
is also known to take on unusual forms that may be in symbiosis with a
central mass concentration, or the larger structure of the galaxy
(i.e., bulge or bar).  A rich phenomenology is seen in the central few
parsecs of galaxy nuclei within the Local Group, but our knowledge of
nuclei at greater distances is limited by a corresponding reduction in
spatial resolution. This problem has particularly afflicted
ground-based optical spectroscopic studies, that typically sample
nuclei with apertures encompassing hundreds of parsecs or more.

To investigate the structure and constituents of galaxy nuclei on
small scales, we therefore undertook a Survey of Nearby Nuclei with
STIS (SUNNS), a program of {\em Hubble Space Telescope} ({\em HST}\/),
narrow-slit spectroscopy for a sample of nearby galaxies outside
the Local Group.  The target
objects were selected on the basis of nebular emission in their
centers, and are thus ``active'' in some sense, although at low
levels that are very common in bright galaxies.  The SUNNS data are
appropriate for addressing several major questions, as described in
part in our previous papers in this series.  The long-slit
measurements of gas kinematics can be used to measure or place limits
on the mass of a central black hole (Sarzi et al. 2001, 2002). The
spectra coupled with {\em HST} imaging can be used further to probe
the structure of nuclear gas disks, which we find to be influenced in
many cases by nongravitational effects (Ho et al. 2002).  The use of
small apertures is very helpful in isolating nuclear emission
components that are lost in the contamination by circumnuclear
sources, as demonstrated by our discovery of unusual broad-line
emission in two galaxies (Shields et al. 2000; Ho et al. 2000).  Small
apertures are also important for obtaining a description of the
stellar population and star formation history within the central tens
of parsecs (Sarzi et al. 2005).

In the current paper we present a summary description of the SUNNS
experimental design, and focus on new results for the nuclear
emission-line characteristics of the sample.  The spatial distribution
and excitation properties of nebular emission can be used to draw
conclusions about the dominant energy sources in galaxy centers, and
as part of our investigation we explore the significance of aperture
effects in determining the spectroscopic appearance of galaxy nuclei.
Understanding the energetics of emission-line nuclei is of fundamental
importance for tracing rates of star formation and black hole growth
in these low-power but nearly ubiquitous objects.  The data reported
here also provide an improved basis for comparing energetic phenomena
and structure in the  Milky Way's nucleus with the centers of other
similar galaxies.

\section{Observations and Measurements}

{\em HST} observations (GO-7361) were obtained for 24 galaxies within
17 Mpc of the Milky Way, drawn from the Palomar spectroscopic survey
of nearby galaxies (Filippenko \& Sargent 1985; Ho, Filippenko \&
Sargent 1997a).  The sample was chosen to be distance-limited to
maximize spatial resolution, and was selected to have nebular emission
potentially suitable for kinematic studies and energetic probes.  We
focused our study on disk galaxies, since these objects most commonly
show nebular emission in their centers.  We restricted the sample to
Hubble types S0 -- Sbc, since the nuclei of later types are more
likely to be weakly defined, affected by dust, or otherwise subject to
confusion.  The sample was limited to galaxies with inclinations $i <
60^\circ$ to minimize extinction by dust residing in the disk.  We
selected objects with H$\alpha$ or [\ion{N}{2}] $\lambda$6583 line
emission $\gtrsim 10^{-15}$ ergs s$^{-1}$ cm$^{-2}$ within a $2\arcsec
\times 4\arcsec$ aperture (from Ho et al. 1997a), with spectroscopic
classifications fairly evenly distributed among Seyfert, LINER 1,
LINER 2, LINER/\ion{H}{2} ``transition,'' and \ion{H}{2} categories.  
The result is listed in Table 1.

Data were acquired in 1998 and 1999 with the Space Telescope Imaging
Spectrograph (STIS) on board {\em HST}, using the 0\farcs 2 $\times$
50\arcsec\ aperture.  For each galaxy, acquisition images were
obtained using the F28X50LP long-pass filter with 30 s integration,
and peak-up exposures were taken in order to center the spectrograph
slit on the target nucleus.  Inspection of these and other archival
{\em HST} images indicates that in all but one case, the aperture was
well-centered on an obvious brightness peak at the morphological
center, with no indication of confusion by dust obscuration or other
structure; errors in aperture centering due to dust effects are
$\lesssim 0.25$ pixels $\approx 0\farcs 012$. The exception occurred
in NGC~4138, where the peak-up procedure positioned the slit on a
bright foreground star offset $\sim 3\farcs 8$ from the nucleus;  we
consequently dropped this object from our analysis.  The slit position
angle (PA) was left unconstrained and thus was set by the spacecraft
roll angle at the time of observation.  For a given galaxy, all
exposures were obtained at a single PA.

Two-dimensional (2-D) spectra were obtained with the G430L and G750M
gratings, yielding spectra spanning 3300--5700 and 6300--6850 \AA\
with FWHM spectral resolution for extended sources of 10.9 and 2.2 \AA
, respectively.  These intervals were chosen in order to include
prominent emission lines, and to measure continuum intervals
diagnostic of stellar populations and possible active galactic nucleus
(AGN) contributions.  The medium-resolution grating was selected for
the red setting so as to enable high precision velocity measurements
of the H$\alpha$ $\lambda$6563 and [\ion{N}{2}] $\lambda\lambda$6548,
6583 lines, which tend to be the strongest optical emission features,
and hence optimal for kinematic studies.  Two exposures totaling $\sim
1800$ s and three exposures totaling $\sim 2700$ s were obtained with
the blue and red settings, respectively.  For most of the target
galaxies, the telescope pointing was adjusted between duplicate
exposures so as to shift the location of the nucleus along the slit by
increments of several pixels, to facilitate removal of hot pixels.

The resulting data were calibrated using standard methods.  The
2-D spectra were bias- and dark-subtracted, flat-fielded, aligned,
and combined into single frames. Cosmic rays and hot pixels remaining
in the combined 2-D spectra were cleaned according to the prescription
of Rudnick, Rix, \& Kennicutt (2000), and the spectra
were then corrected for geometrical distortion and wavelength and flux
calibrated with standard STSDAS procedures within
IRAF\footnote[11]{IRAF is distributed by the National
Optical Astronomical Observatory, which is operated by AURA,
Inc. under contract to the NSF.}. During each reduction step, care was
taken to correctly propagate the information contained in the error
array corresponding to each galaxy 2-D spectrum and initially provided
by the STIS pipeline.

To represent the nuclear spectrum of each galaxy, we extracted
aperture spectra five pixels wide ($\sim 0\farcs25$), centered on the
brightest part of the continuum, using the IRAF task
{\scriptsize APALL}. The extracted spectra thus
consist of the central emission convolved with the STIS spatial
point-spread function and sampled over an aperture of $0\farcs25
\times 0\farcs2$, roughly corresponding to a circular aperture with
radius $R = 0\farcs13$, or 8.2 pc for the mean sample distance of
$\sim$13 Mpc (or $R = 4 - 11$ pc for the full range of galaxy distances).  
In addition, from the 2-D error array corresponding to
each 2-D spectrum, we extracted the central five, one-pixel-wide
aperture spectra, and added them quadratically to obtain the error
array corresponding to each nuclear extraction.  The resulting 1-D
spectra for the blue (G430L) setting appear in Sarzi et al. (2005),
while those for the red (G750M) setting are shown here in Figure 1.

\begin{figure}
\figurenum{1}
\plotone{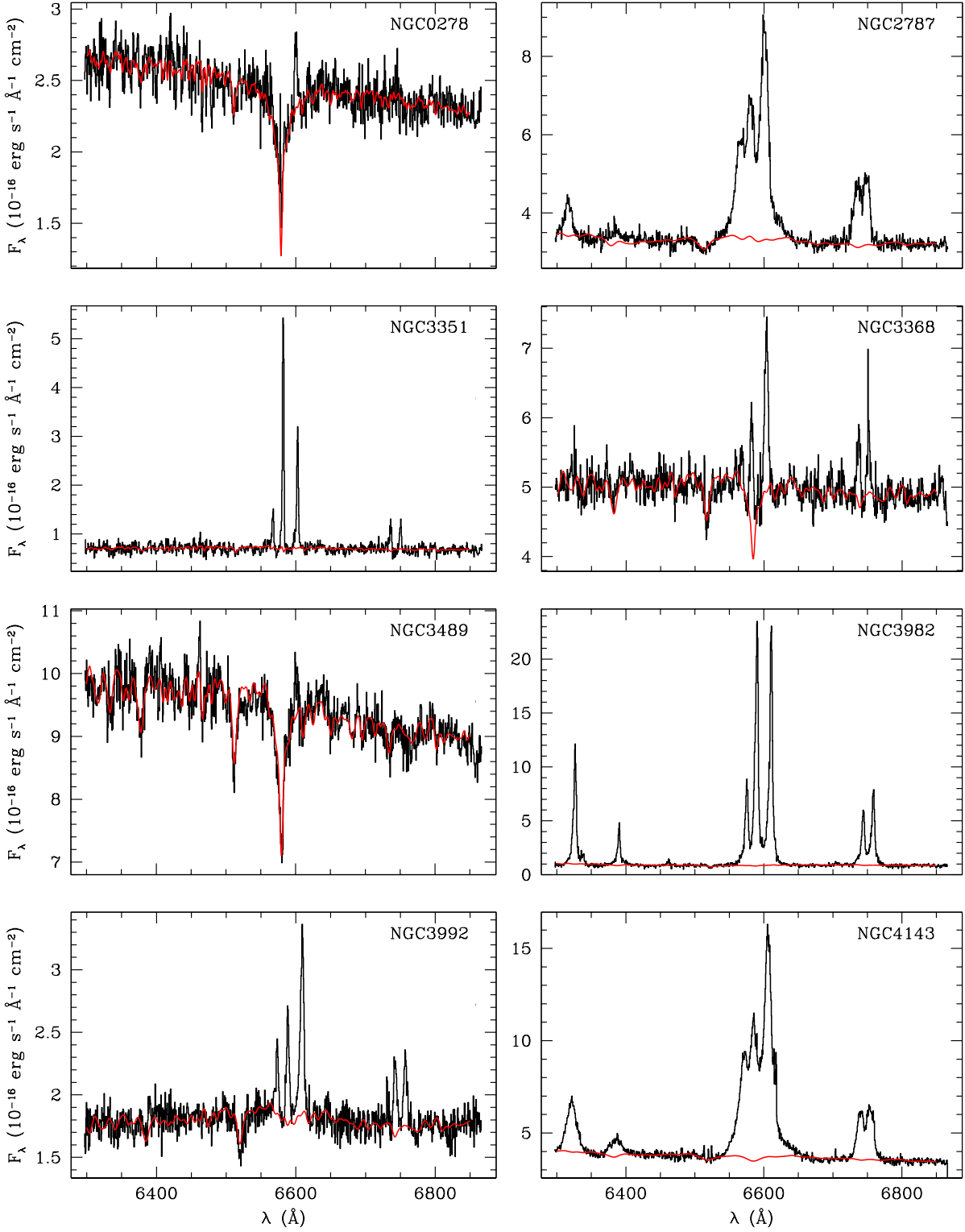}
\caption{Spectra of the sample galaxy nuclei acquired with the G750M
grating, as a function of observed wavelength (black), with fits
to the stellar continuum overplotted (red).}
\end{figure}
\begin{figure}
\figurenum{1}
\plotone{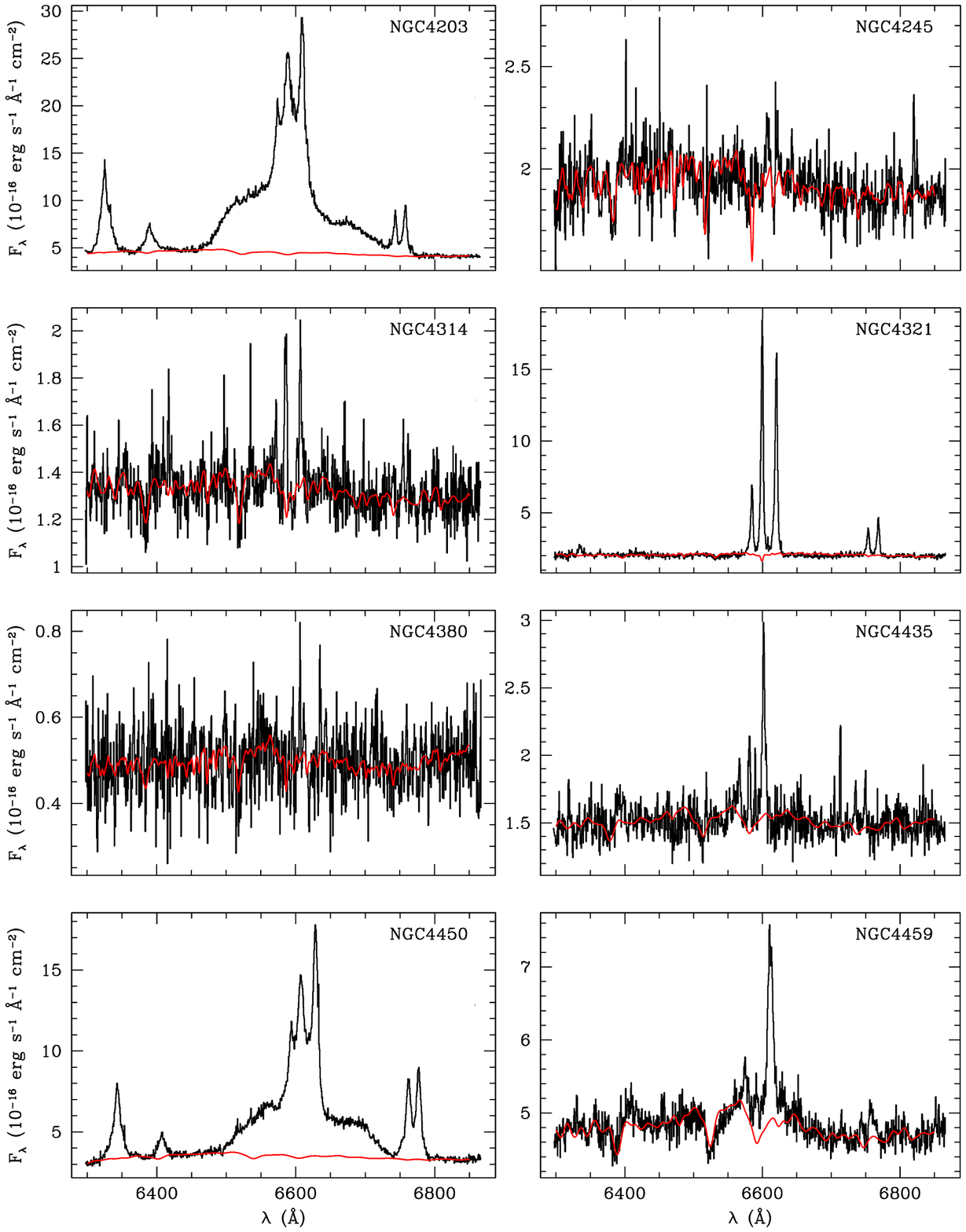}
\end{figure}
\begin{figure}
\figurenum{1}
\plotone{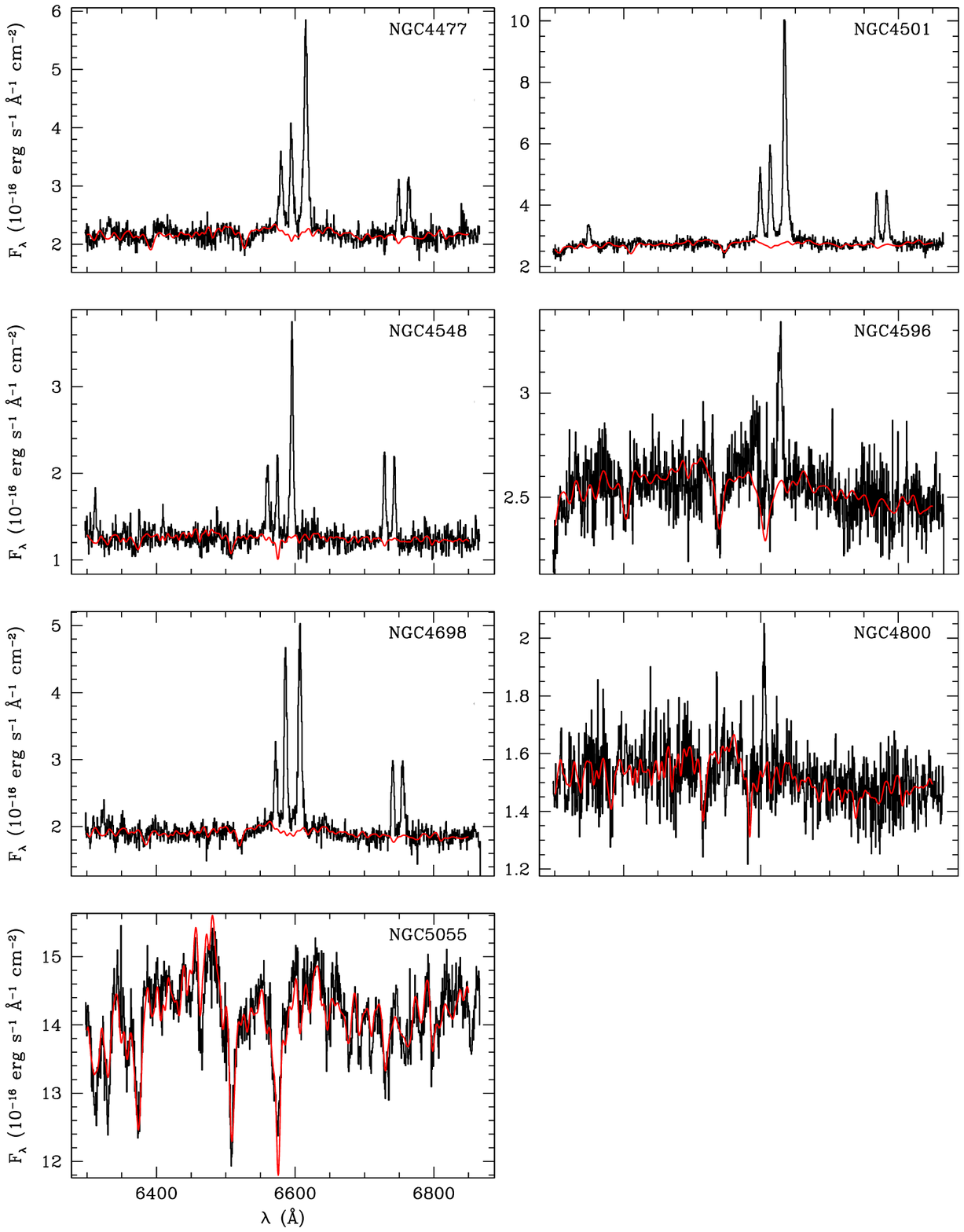}
\end{figure}

The emission-line equivalent widths in the nuclear spectra are often
small and consequently require subtraction of the underlying starlight
in order to measure accurate emission-line fluxes.  
We modeled the continuum with linear combinations of stellar spectra
from the MILES library of Sanchez-Blazquez et al. (2006), with weights
and velocity broadening optimized with the direct-fitting method
described in Sarzi et al. (2006).
This is similar to the method of Rix \& White (1992), except that
the spectral regions affected by nebular emission are no longer
excluded from the fitting process, since the emission-lines are
treated as Gaussian templates and fitted simultaneously with the
stellar templates to the observed spectra.
This has the advantage of maximising the spectral information available to
the fitting algorithm.

As the blue and red spectra have different spectral resolutions they
cannot be fitted simultaneously. On the other hand, the presence of
stronger emission lines and the limited number of absorption features
makes it difficult to independently match the stellar continuum of the
red bandpass. 
We therefore fit first the blue spectra, which are important for
constraining the mix of template stellar types, and use the resulting
optimal combination of templates to match the red spectra, allowing
only for a different velocity broadening.
The fitting of the blue spectra include reddening due to interstellar
dust, both in the Milky Way and in the sample galaxies, and due to
dust in the emission-line regions. 
The latter affects only the fluxes of the emission-line templates, and
is constrained by the expected decrement of the Balmer lines. 
In matching the blue spectra we imposed the same kinematics for both
forbidden and recombination lines. 
We also restricted the stellar template library to spectra with values
for the H$\beta$ absorption line-strength index exceeding 0.1\AA.

In the present study we are interested in obtaining the best empirical
fit to the starlight, and no a priori assumptions are made concerning
the relative weighting of different stellar types; our method here
consequently differs somewhat from that employed in our population
analysis (Sarzi et al. 2005) which is based on templates corresponding
to parametrized star formation histories. 
The fits to the blue spectra are very similar to the parametrized fits
plotted in Sarzi et al (2005), although the MILES spectra can 
match several spectral regions where models such as those of Bruzual \&
Charlot (2003) would need to account for non-solar abundance of
$\alpha$-elements before they can match the data.  

The best description for the stellar continuum in the red bandpass
is overplotted on the spectra shown in Figure~1. 
The fits for the red spectra differ from those for the blue bandpass
not only in that they use a different velocity broadening and the
optimal combination of templates from the fit on the blue spectra, but
also in that they employ a generalized multiplicative second-order polynomial
adjustment instead of a specific dust reddening law. Emission-line
templates were used also while fitting the red spectra, although for
galaxies with complex line profiles or broad components, the regions
affected by emission were simply excluded from the fit.

For many of our target objects, the emission lines within the nuclear
aperture are weak.  We attempted to measure lines that were visible in
much of the sample on inspection, or likely to be present based on the
behavior of line-intensity ratios in other nebular
sources. Specifically, we measured or placed limits on the flux in
[\ion{O}{2}]~$\lambda$3727, H$\gamma$~$\lambda$4340,
H$\beta$~$\lambda$4861, [\ion{O}{3}]~$\lambda\lambda$4959, 5007,
[\ion{O}{1}]~$\lambda$6300, [\ion{N}{2}]~$\lambda\lambda$6548, 6583,
H$\alpha$~$\lambda$6563, and [\ion{S}{2}]~$\lambda\lambda$6716, 6731.
The lines were measured from the starlight-subtracted spectra using
{\scriptsize SPECFIT} (Kriss 1994) as implemented in IRAF.  This
routine is suitable for fitting complicated line profiles
with additional contraints between lines, while also generating error
estimates for measured quantities.  Gaussian
profiles were assumed for the lines; a few sources required double
Gaussians to obtain a good match to the observed lines, and additional
components were included in cases where there was evidence of broad
Balmer emission (\S 3.3).

To improve the fitting accuracy, we
introduced constraints to reduce the number of fit parameters when
justified by physical arguments, or by our previous experience with
ground-based spectra (e.g., Ho et al. 1997a).  The H$\alpha$,
[\ion{N}{2}], and [\ion{S}{2}] lines were assumed to share a common
centroid velocity, and the [\ion{N}{2}] and [\ion{S}{2}] lines were
assumed to share a common velocity width.  A flux ratio of 1:3 was
assumed for the [\ion{N}{2}] $\lambda\lambda$6548, 6583 and
[\ion{O}{3}] $\lambda\lambda$4959, 5007 doublets, as dictated by
atomic physics (e.g., Osterbrock 1989).  The H$\beta$ and H$\gamma$
lines were also assumed to share a common velocity and profile.  The
G430L and G750M spectra were fit separately because of their differing
spectral resolution and possible zero-point wavelength offsets.  In
determining relative wavelengths for lines with shared velocities,
rest wavelengths to 6 significant figures were taken from Kaufman \&
Sugar (1986) for the forbidden lines and from Wiese, Smith, \& Glennon
(1966) for the Balmer lines.  The resulting line fluxes are listed in
Table 2.

The line fluxes are uncertain due to both statistical fluctuations and
systematic errors.  {\scriptsize SPECFIT} uses a $\chi^2$ minimization
algorithm, which provides one estimate of the line flux uncertainties.
As a measure of the possible systematic error, we also calculated an
equivalent 1$\sigma$ Gaussian uncertainty based on the standard
deviation $\sigma_c$ per pixel in the residual continuum after
starlight subtraction.  The standard deviation was computed in regions
devoid of emission lines and traces structure due to imperfect
continuum subtraction, in addition to statistical fluctuations.  The
corresponding error is then given by a Gaussian profile with amplitude
of 1$\sigma_c$, and a width typical of the measured narrow lines (FWHM
of 300 km s$^{-1}$ for G750M and 550 km s$^{-1}$ for G430L, where the
width includes instrumental broadening).  In determining a final value
for the flux uncertainty, we adopted the largest of 1) the
{\scriptsize SPECFIT} error, 2) the Gaussian 1$\sigma$ uncertainty,
and 3) a conservative minimum error of 10\% of the line flux.  As a
consistency test, we repeated the line measurements after subtracting the
best-fit multi-age stellar population models for these spectra that are
described by Sarzi et al. (2005).  Line fluxes were measured from the
continuum-subtracted spectra by direct integration over the line profile,
using {\scriptsize SPLOT} in IRAF.  The results showed good consistency
within the measurement uncertainties with the values listed in Table 2.

\section{Results}

The {\sl HST} aperture employed for this study is much smaller
than typical ground-based apertures, and a comparison of nebular
properties as seen on these different scales provides useful
information on the structure of emission nuclei.  The Palomar spectra
used to define the current sample employed a $2\arcsec \times
4\arcsec$ aperture, which is $\sim 10 \times$ larger in linear
dimension and $160 \times$ larger in area than the SUNNS nuclear
aperture.  The linear scales probed by these data are $\sim 100$ vs
$\sim 10$ pc.

\subsection{Spatial Extent of Emission}

The degree to which the nebular emission in our sample galaxies is
centrally concentrated can be gauged from a comparison of fluxes
measured through the SUNNS and Palomar apertures.  The result is shown
in Figure 2, which displays the ratio of narrow H$\alpha$ fluxes as a
function of Palomar spectroscopic class.  The sequence of categories
on the abscissa is intended to reflect the degree to which the nuclei
show evidence of being powered by accretion.  In this and subsequent
figures displaying ratios, the plotted error bars are propagated from
the uncertainties listed in Table 2 for the SUNNS measurements and in
Ho et al. (1997a) for the Palomar data.  For a uniform nebular
surface brightness, the flux ratio should be $-2.2$ dex.  In the majority
of cases the ratio is measured to be higher, indicating some degree
of central concentration.  On the other hand, for all of the galaxies,
the small-aperture measurement shows a significant reduction in flux
relative to the Palomar value, indicating that the central nebulae in
these objects are resolved at {\sl HST} resolution.

\begin{figure}
\figurenum{2}
\epsscale{0.6}
\plotone{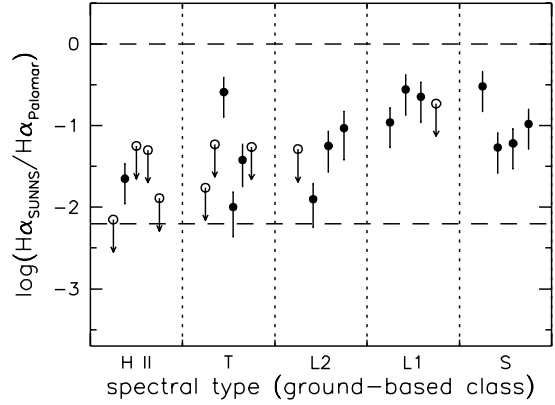}
\caption{Ratio of narrow H$\alpha$ flux measured in the SUNNS data relative
to that measured at Palomar, as a function of spectroscopic class as
determined from the Palomar spectra (T = \ion{H}{2}/LINER transition 
source, L2 = LINER 2, L1 = LINER 1.9, S = Seyfert 1.9 or 2).  Upper
limits are represented by open symbols.  The upper dashed line represents
a unit flux ratio, and the lower dashed line represents the value
predicted from the ratio of aperture areas, for a source with uniform
surface brightness.}  
\end{figure}

Figure 2 shows evidence of a trend in that sources showing the
strongest indications of accretion power (LINER 1s and Seyferts) display
the largest ratios, implying greater concentration.  In contrast, the
\ion{H}{2} nuclei show the smallest ratios, and in fact four of the
five \ion{H}{2} nuclei are undetected in H$\alpha$ in the SUNNS
aperture.  The star formation powering \ion{H}{2} ``nuclei''
apparently resides on scales of tens of parsecs or more (see, e.g.,
Hughes et al. 2003, Cid Fernandes et al. 2004, and Gonz\'alez Delgado 
et al. 2004 for related results).  The narrow
emission-line equivalent widths (EWs; see Table 2 and Figure 3) show a
similar trend\footnote{Error bars in Figure 3 are smaller in some
cases than in Figure 2 because the equivalent widths are insensitive to
absolute flux calibration uncertainties in the Palomar spectra.}, 
with larger EW(H$\alpha$) found in the SUNNS
aperture for the LINER 1 and Seyfert nuclei, but smaller EW(H$\alpha$)
in the small aperture for the majority of \ion{H}{2}, transition, and
LINER 2 nuclei.

\begin{figure}
\figurenum{3}
\epsscale{0.6}
\plotone{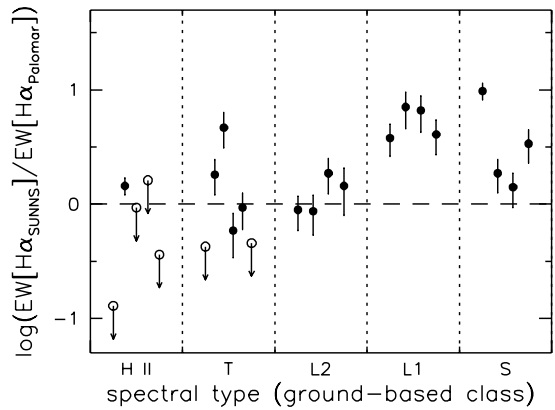}
\caption{Ratio of narrow H$\alpha$ equivalent width measured in the
SUNNS data relative to that measured at Palomar (Ho et al. 1997a), as a
function of spectroscopic class as determined from the Palomar spectra.
Upper limits are represented by open symbols.}
\end{figure}

\subsection{Emission Line Ratios}

While a direct measurement of line ratio gradients is possible in
principle from the {\sl HST} long-slit data, in practice the
signal-to-noise ratio severely hampers detailed study of spatial
variations.  Comparison of line ratios measured in the SUNNS and
Palomar apertures provides an alternative means of quantifying
excitation gradients.  In addition to yielding a factor of $\sim 10 
\times$ more signal, the Palomar aperture avoids possible problems
with azimuthal variations in extended emission that are not adequately
sampled by the narrow SUNNS slit.
Line ratios useful as diagnostics that are also measured for a
significant number of sample members are limited, but four
particularly relevant ratios are [\ion{O}{3}] $\lambda$5007/H$\beta$,
[\ion{O}{1}] $\lambda$6300/H$\alpha$,
[\ion{N}{2}]$\lambda$6583/H$\alpha$, and [\ion{S}{2}]
$\lambda\lambda$6716, 6731/H$\alpha$.  These diagnostics are commonly
employed in classification of nebular sources (e.g., Veilleux \&
Osterbrock 1987), and standard diagrams used for this purpose
are shown in Figure 4, with measurements from both Palomar and SUNNS
plotted.

\begin{figure}
\figurenum{4}
\epsscale{1}
\plotone{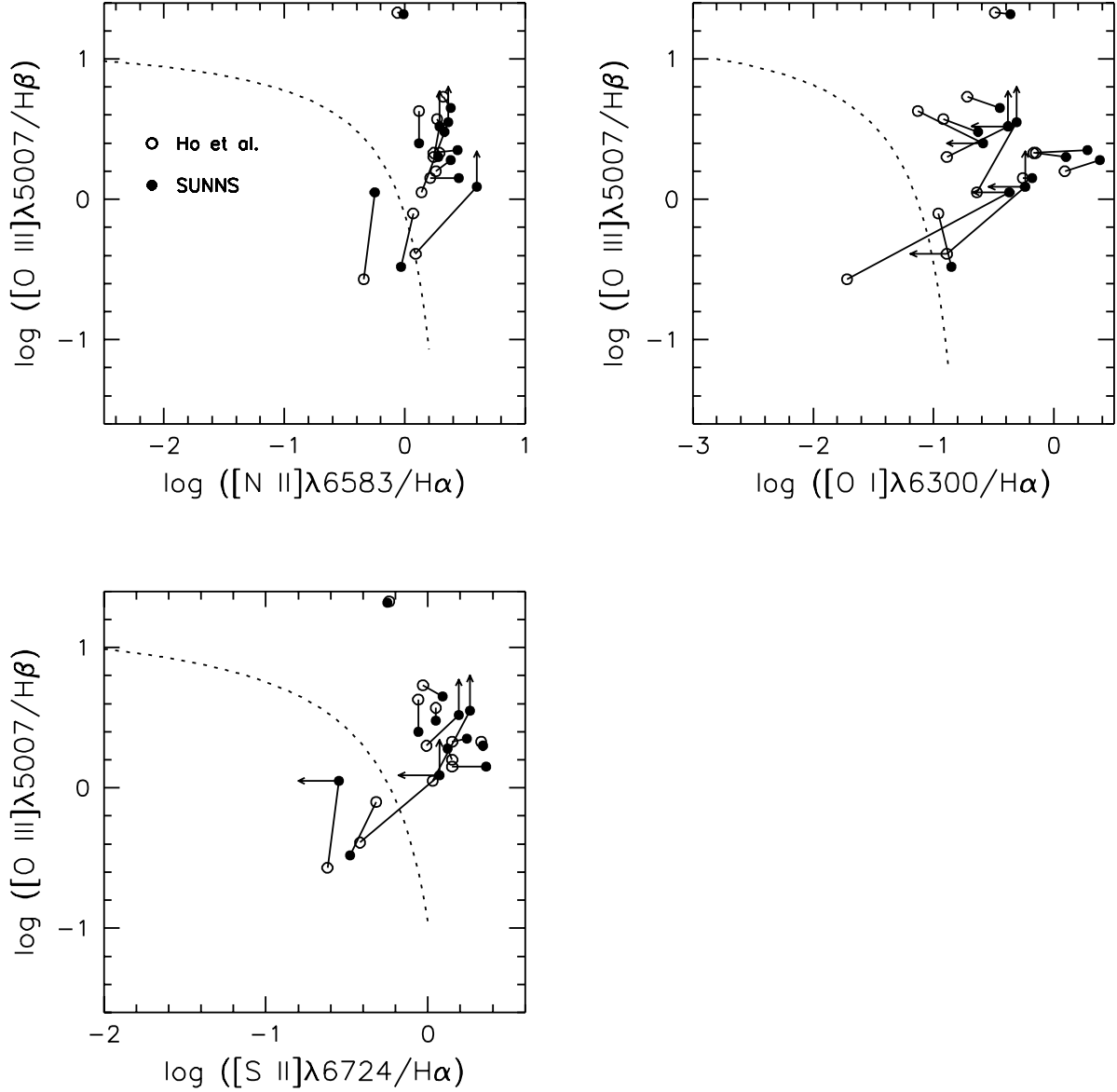}
\caption{Emission line intensity ratios as measured through the large
Palomar aperture (Ho et al. 1997a) and by SUNNS.  The dotted line is
the boundary between starburst and non-starburst sources given by 
Kewley et al. (2001).  [\ion{S}{2}] $\lambda$6724 represents the sum
of [\ion{S}{2}] $\lambda\lambda$6716, 6731.}
\end{figure}

There is considerable variation in the behavior of different
objects in Figure 4, but several conclusions can nonetheless be drawn
from these plots.  First, the line ratios are often rather insensitive
to aperture size, and in the present comparison show changes that are
usually less than a factor of two.  Related to this, the sources
generally do not shift their position by large amounts in the line
ratio diagrams, which means that their classification is insensitive
to aperture size.  This finding has interesting and important implications
that we examine in detail below.  One limitation of the two-dimensional
line ratio diagrams is that the SUNNS observations provide only upper
limits for both [\ion{O}{3}] and H$\beta$ for nearly half the sample,
thus preventing their inclusion in Figure 4.  An alternative is to 
compare line intensity ratios individually, and for this purpose we plot
in Figure 5 the ``ratio of ratios'' for the measurements obtained in
the two apertures.

\begin{figure}
\figurenum{5}
\epsscale{0.5}
\plotone{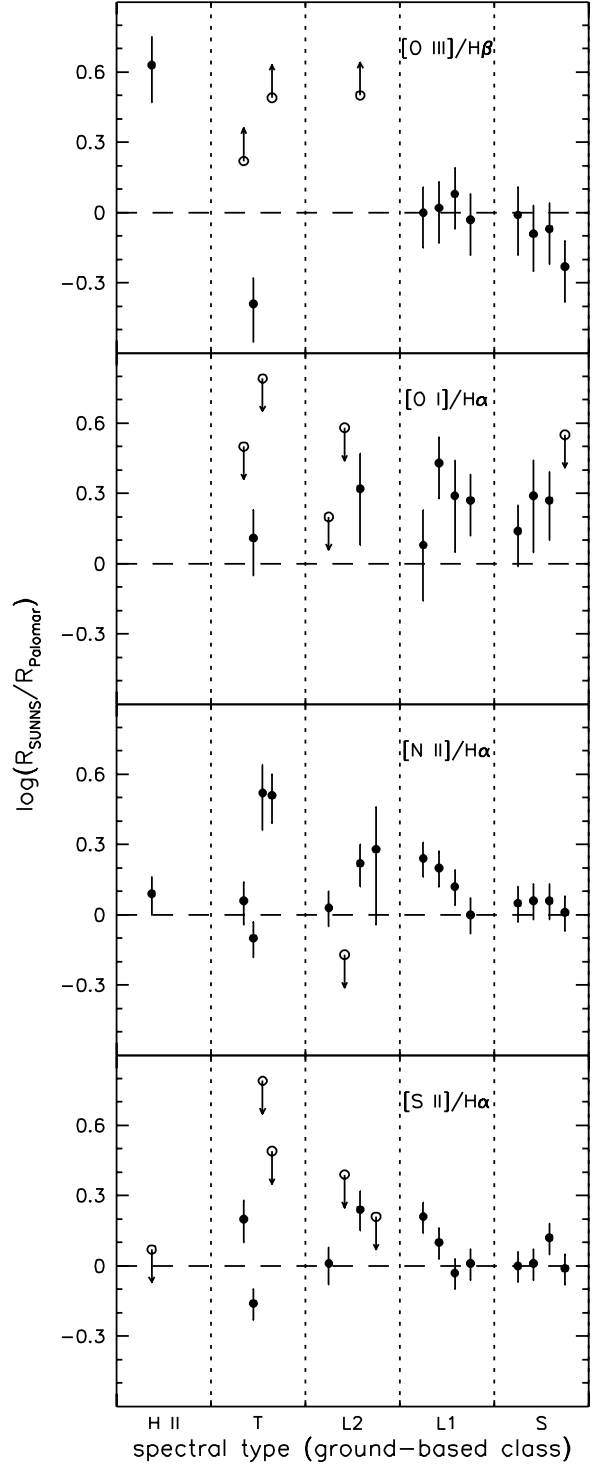}
\caption{SUNNS/Palomar ratios of the indicated diagnostic flux ratios,
as a function of spectroscopic class.  Limits are represented by 
open symbols. Points for a given source have a common horizontal 
location (offset for clarity) within the classification bins in 
all plots shown in Figures 2, 3, and 5.}
\end{figure}

Several patterns are discernable in Figure 5.  [\ion{O}{1}]/H$\alpha$
is arguably the best indicator of excitation beyond the contributions
of normal \ion{H}{2} regions, since it requires a significant
partially ionized zone for its production.  The [\ion{O}{1}] line is
in fact most significant in the LINER 1 and Seyfert nuclei, and Figure
5 shows a clear tendency for the ratio to be larger in the small
aperture for these sources.  A weaker trend in the same direction 
exists for [\ion{N}{2}]/H$\alpha$, but no trend is evident for
[\ion{S}{2}]/H$\alpha$.  The more limited measurements
of [\ion{O}{3}]/H$\beta$ show little dependence on aperture size,
except for some tendency for the ratio to decrease in the small
aperture in the Seyfert nuclei.

Density gradients may be a factor in the nebular emission structure
reflected in Figure 5. Previous studies have shown indirect evidence
from line width-critical density correlations (e.g., Pelat, Alloin, \&
Fosbury 1981; Filippenko \& Halpern 1984) as well as direct evidence from
the [\ion{S}{2}] density diagnostic (Barth et al. 2001) that densities
increase toward small radii in the narrow-line regions of AGNs.
Figure 6 shows the density-sensitive [\ion{S}{2}]
$\lambda$6716/$\lambda$6731 line ratio for the sample sources as
measured in both the Palomar and SUNNS spectra.  The ratio is lower in
all but one case for the small aperture measurement, implicating a
characteristically higher density on smaller radial scales; median
values are 1.2 from the Palomar measurements and 0.95 from SUNNS,
corresponding to densities of $\sim 200$ cm$^{-3}$ and $\sim 700$
cm$^{-3}$, respectively.  These values are presumably
emissivity-weighted averages of a potentially much broader span of
densities.  The different spectral classes show substantial overlap in
their distributions in Figure 6.

\begin{figure}
\figurenum{6}
\epsscale{0.6}
\plotone{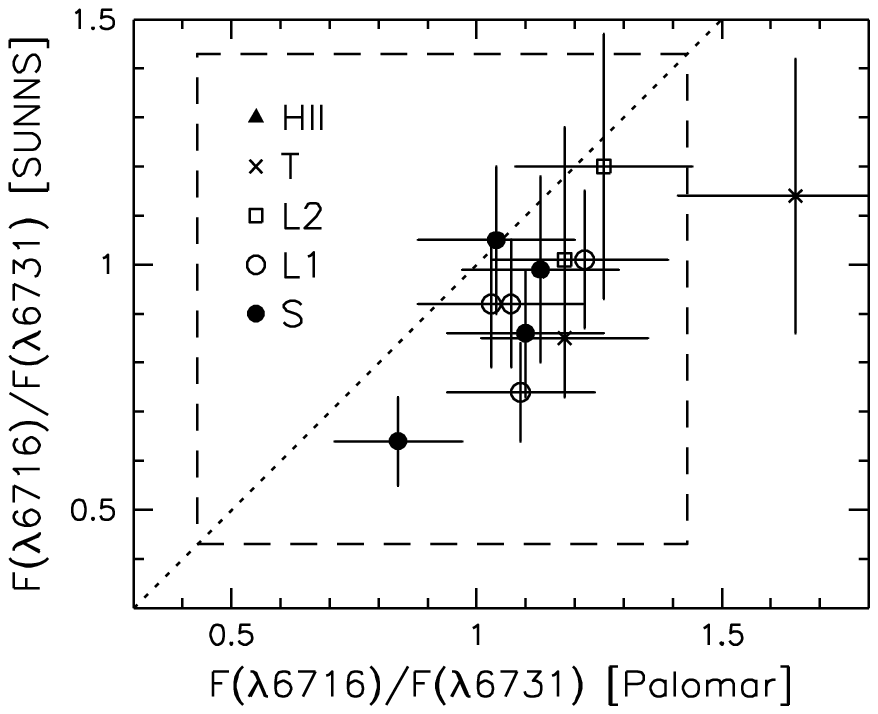}
\caption{Line flux ratio [\ion{S}{2}]$\lambda$6716/[\ion{S}{2}]$\lambda$6731
as measured by SUNNS versus Palomar.  The dotted line represents $x = y$
and the dashed box outlines the boundary of allowed values predicted
from theory, defined by the high- and low-density limits.  The data
reveal a strong trend of lower ratios, and hence higher average densities, in
the SUNNS aperture.}
\end{figure}

The different behavior observed in the [\ion{O}{1}], [\ion{N}{2}], and
[\ion{S}{2}] lines in Figure 5 can potentially be understood as a
consequence of the varied critical densities for the forbidden lines.
In the SUNNS aperture a significant fraction of the nebular gas for
the LINERs and Seyferts evidently has a density similar to or greater
than the [\ion{S}{2}] $\lambda\lambda$6716, 6731 critical densities
$n_{crit}$ of $\sim 1.5 \times 10^3$ cm$^{-3}$ and $\sim 3.9 \times
10^3$ cm$^{-3}$ (e.g., Peterson 1997), which could act to depress the
[\ion{S}{2}]/H$\alpha$ ratio in the small aperture.  The [\ion{S}{2}]
$n_{crit}$ values are the lowest of those for the lines shown in
Figure 5.  In contrast, [\ion{O}{1}] $\lambda$6300, with $n_{crit}
\approx 1.8 \times 10^6$ cm$^{-3}$, remains strong in the small
aperture; the larger [\ion{O}{1}]/H$\alpha$ ratios found with SUNNS
may reflect an increased importance of partially ionized gas on small
scales, but would also be consistent with a higher nebular temperature
in dense regions where cooling by low-$n_{crit}$ emission (e.g., from
infrared fine-structure lines) is inefficient.
[\ion{N}{2}] $\lambda$6583, with an intermediate $n_{crit}$ value of
$\sim 8.7 \times 10^4$ cm$^{-3}$, shows a weaker trend in the same
direction.  The overall results suggest that a significant fraction of
the nebular gas in galaxies on scales of $\sim 10$ pc has densities of
$\sim 10^4$ cm$^{-3}$.  The [\ion{O}{3}] $\lambda$5007 line, with
$n_{crit} \approx 7 \times 10^5$ cm$^{-3}$, does not readily fit into
this scenario.  For the Seyfert nuclei in particular, the 
[\ion{O}{3}]/H$\beta$ ratio tends
to decrease in the small aperture, and there is some degree of
negative correlation between the [\ion{O}{3}]/H$\beta$ and
[\ion{O}{1}]/H$\alpha$ ratios, suggesting that ionization structure --
which will be sensitive to ionization parameter and hence the detailed 
radial variation in density -- also plays a role in the aperture 
dependence.

\subsection{Broad Lines}

The existing Palomar observations provided evidence that six of our 23
target galaxies exhibit weak quasar-like broad H$\alpha$ emission,
resulting in classification as LINER 1.9 (NGC~2787, NGC~4143,
NGC~4203, NGC~4450) or Seyfert 1.9 (NGC~3982, NGC 4501) nuclei (Ho et
al. 1997b).  We investigated whether a broad component was also
present in the {\sl HST} spectra for these sources and the other
objects in our sample.  The SUNNS data for two of the LINER 1.9
galaxies, NGC~4203 and NGC~4450, display evidence of not only a
Gaussian broad H$\alpha$ line, as seen from the ground, but also a
very broad, double-shouldered component (see Fig. 1); these findings
are discussed in detail elsewhere (Shields et al. 2000; Ho et
al. 2000).  We attribute the discovery of the additional broad
features to the improved contrast in the small {\sl HST} aperture
between the broad emission, which presumably arises within $\la 1$ pc
of the center, and the stellar continuum plus narrow-line emission,
which is more widely dispersed in origin.  The implication is that
such broad emission may be common in low-luminosity AGNs, and a
signature of the accretion structure in low accretion-rate nuclei.
Other examples of double-shouldered emission profiles found in
{\em HST} spectra of LINER 1 nuclei are described by Bower et
al. (1996) and Barth et al.  (2001).

Figure 7 shows the results of a profile decomposition using the SUNNS
data for the remaining four galaxies where broad emission was
previously suspected.  A common profile shape was employed to represent the
[\ion{N}{2}], narrow H$\alpha$, and [\ion{S}{2}] lines; double
Gaussians were adopted to better match the profiles for NGC~3982 and
NGC~4143.  An additional broad H$\alpha$ component with unconstrained
width and central wavelength was included in the fit.  Strong evidence
for a broad emission feature is found in NGC~2787 and NGC~4143, while
marginal evidence for such emission is present in NGC~4501, and a
reasonable fit for NGC~3982 can be obtained with no broad component.
The narrow-line fluxes resulting from this decomposition
are listed in Table 2;  broad-line fluxes and the fraction $f_{blend}$
of the total H$\alpha$+[\ion{N}{2}] complex represented by the broad
H$\alpha$ feature are listed in Table 3, along with the corresponding
numbers measured in the Palomar spectra.  For the four galaxies where
broad emission appears to be confirmed (NGC~2787, NGC~4143, NGC~4203,
and NGC~4450), $f_{blend}$ increases if these values are taken at face
value, as expected if the narrow emission is resolved and diminished
in the small aperture while the broad emission remains unaffected by
slit losses.  

\begin{figure}
\figurenum{7}
\epsscale{0.9}
\plotone{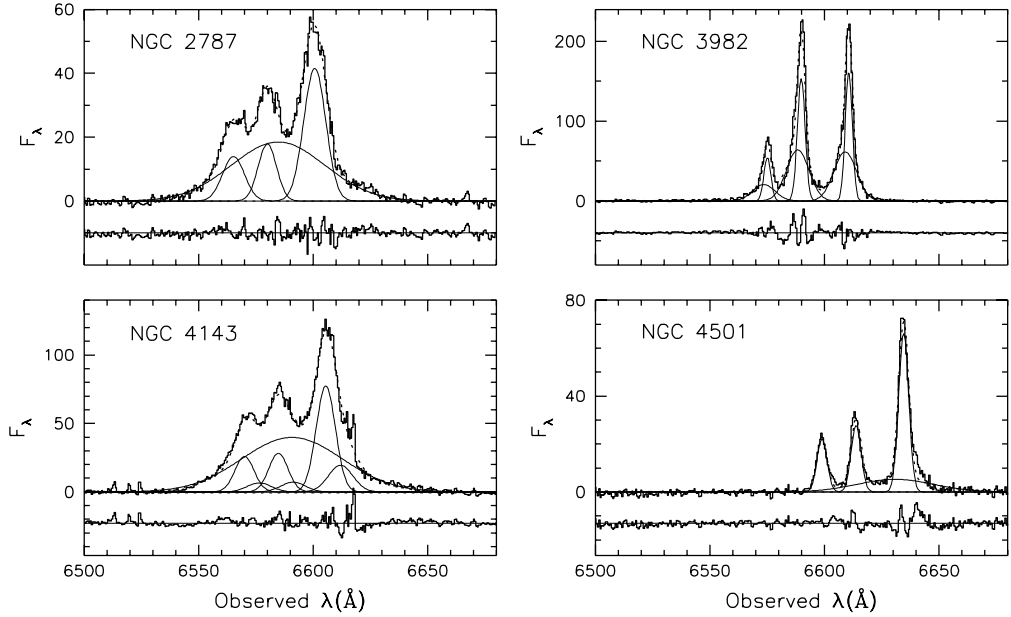}
\caption{Continuum-subtracted spectra in the H$\alpha$ + [\ion{N}{2}]
region, plotted with Gaussian model components (solid curves), the
reconstructed blend (dotted line), and residuals (offset vertically
for better visibility).}
\end{figure}

Closer scrutiny of the deblending results suggests a problem with the
fitting procedure, however.  The values listed in Table 2 imply
H$\alpha$/H$\beta$ ratios for the narrow component of ($0.9\pm 0.1$,
$1.0\pm 0.1$, $0.50\pm 0.07$, $1.0 \pm 0.1$) for NGC~2787, NGC~4143,
NGC~4203, and NGC~4450, respectively.  These ratios are considerably
smaller than the theoretical ratio of 3.1 predicted for AGN
narrow-line regions (Osterbrock 1989), and are impossible to
understand on physical grounds.  The low values are unlikely to be caused by
broad-line contamination of H$\beta$; NGC~4203 and NGC~4450 both show
evidence of broad emission in H$\beta$ and H$\gamma$, but this
component is included as a separate component when determining the
narrow Balmer fluxes.  No evidence of broad emission is seen in these
lines for NGC~2787 and NGC~4143.  The narrow H$\gamma$/H$\beta$ ratios
are, moreover, normal: NGC~2787, NGC~4143, NGC~4203, and NGC~4450
display H$\gamma$/H$\beta$ = $0.53\pm 0.07$, $<0.43$, $0.50\pm
0.07$, and $0.55\pm 0.08$, respectively, in reasonable agreement with
the predicted recombination value of $\sim 0.47$.  There is no
evidence of significant discontinuities between the G430L and G750M
spectra indicative of flux calibration errors; almost all of the other objects in
our sample show narrow H$\alpha$/H$\beta$ ratios consistent with the
theoretical prediction (NGC~4698 has a ratio of $2.5 \pm 0.4$), or a 
larger ratio attributable to reddening.
It is noteworthy that for the four galaxies in question, the forbidden
line/H$\alpha$ flux ratios are unusually large for LINERs or Seyfert
nuclei (although it must be kept in mind that the SUNNS aperture does
not sample the full narrow-line region).

Given these findings, we strongly suspect that the deblending
procedure outlined above and illustrated in Figure 7 underestimates
the true narrow H$\alpha$ flux which in turn would produce an
overestimate of the broad component flux.  We experimented with
alternative parameterizations of the fits for NGC~2787 and NGC~4143,
including decoupling of the narrow H$\alpha$ line widths from the
other lines, forcing the narrow H$\alpha$ flux to match the value
predicted from H$\beta$, and forcing the narrow H$\alpha$ width to
match that of H$\beta$.  In general, these approaches were not
particularly satisfactory at matching the detailed profile structure,
resulting in residuals larger than seen in Figure 7.  This outcome
suggests that the narrow and/or broad profiles are rather more complex
than simple Gaussians and that significant differences may exist
between the different narrow-line profiles.  A possible broader
component of [\ion{N}{2}] than is currently represented in our 
deblending model may also be suggested by the redward displacement
of the broad H$\alpha$ feature relative to the narrow emission found
in the four galaxies at issue here, since the stronger line in the
doublet, $\lambda$6583, would thereby add flux that our default
model would interpret as redshifted H$\alpha$ emission.

We therefore caution that
the fluxes of narrow H$\alpha$ and possibly [\ion{N}{2}] in Table 2
for these objects may suffer from systematic errors in excess of the
listed uncertainties.  For purposes of comparison with the Palomar
results, however, the fact that we employed a very similar fitting
algorithm may mitigate the influence of these errors.  It is
noteworthy that the Palomar measurements from Ho et al. (1997a) yield
H$\alpha$/H$\beta$ ratios of 1.9, 1.6, 1.0, and 2.6 for NGC~2787,
NGC~4143, NGC~4203, and NGC~4450, respectively, which likewise fall
short of the recombination prediction and suggest similar systematic
errors to what we have encountered with the SUNNS data.  In ratioing
the SUNNS measurements to the Palomar measurements (e.g., in
Figures 2, 3, and 5 ), we can thus expect
these errors to cancel out at least partially.  A best estimate 
of the true narrow H$\alpha$ flux is probably that derived from the 
narrow H$\beta$ flux, scaled by the recombination prediction, which
yields ($6.2 \pm 0.6$, $12 \pm 1$, $28 \pm 3$, $12 \pm 1$) $\times 
10^{-15}$ ergs s$^{-1}$ cm$^{-2}$ for NGC~2787, NGC~4143, NGC~4203,
and NGC~4450, respectively.

Comparison of the SUNNS broad H$\alpha$ fluxes with those from Palomar
reveals a surprising trend, in that the SUNNS flux is in all cases
smaller.  Taking the profile decomposition results at face value, the
median reduction is a factor of $\sim 3$.  It is not clear how this
outcome could result from systematic errors in the deblending process.
We also do not expect the broad feature to suffer from aperture losses
in the SUNNS data, since the emission is expected to arise from
sub-parsec scales.  For spatially unresolved sources the STIS flux
calibration corrects for aperture transmission losses to within an
accuracy of $\sim 5$\% at H$\alpha$ (Bohlin \& Hartig 1998).  The
present result could be understood physically if the broad emission is
variable; the sources in the Palomar survey identified as L1.9 or S1.9
nuclei would then have preferentially been in a high flux state, since
sensitivity limits dictate that objects in low flux states would not
be detected as broad-line objects.  If the sources vary on timescales
shorter than the decade interval between the Palomar and SUNNS
observations, we would then expect that the broad H$\alpha$ flux would
be lower in a majority of cases in the second-epoch observation.  We
would also predict that some number of sources seen as Type 2 nuclei
in the Palomar data would display new detectable broad components at
the second epoch. Variability in broad line emission would not be
surprising and has been reported in similar low-luminosity AGNs
previously (e.g., Eracleous \& Halpern 2001 and references therein).

For the sources in our sample that appear in the Palomar spectra as
narrow-line objects, it is therefore a surprise that the {\sl HST}
spectra provide no new indications of broad H$\alpha$.  This
conclusion was reached after examining the residuals from fitting with
purely narrow-line profiles as described above.  The lack of newly
detected broad-line sources is curious with respect to the variability
prediction noted above, but also given the expected increase in
sensitivity in the small {\sl HST} aperture due to the reduction in
circumnuclear contamination.  Investigation of a larger sample
would be desirable in order to clarify the generality of this result
and its implications.

\section{Discussion}

\subsection{AGN Classification and Aperture Size}

An interesting question that can be addressed with the results of this
study is the extent to which the spectroscopic classification of a
galaxy nucleus depends on the measurement aperture employed.  The
results from \S 3.1 indicate that the SUNNS aperture typically
excludes most of the emission that is seen in the Palomar measurement,
and that nebulosity that is clearly dominated by an accretion source
tends to be the most centrally concentrated.  We might therefore
expect the {\sl HST} aperture to isolate gas that is higher in
excitation/ionization and more dominated by the central engine, to the
extent that one is present.  Likewise, the small-aperture measurement
should increase the probability of detection for broad H$\alpha$
components.  In summary, the simple prediction from these
considerations is that the SUNNS nuclear spectra should appear more
AGN-like than the corresponding Palomar spectra, for nuclei where
accretion is important as a power source.

Objects of particular interest in this regard are the \ion{H}{2}/LINER
transition nuclei.  These sources are intermediate in their emission
properties between \ion{H}{2} nuclei and LINERs or Seyfert nuclei
(e.g., Ho et al. 1997a; Hao et al. 2005); other labels applied
to such objects are ``composite nuclei'' (e.g., Kennicutt, Keel, \& Blaha 
1989; V\'eron, Gon\c calves, \& V\'eron-Cetty 1997) 
and ``weak [\ion{O}{1}] LINERs'' (e.g., Filippenko
\& Terlevich 1992).  The standard interpretation for these sources,
which are very common ($\sim 25$\% of bright galaxies; Ho, Filippenko,
\& Sargent 1997c), is that their spectra trace a mix of AGN emission and
circumnuclear \ion{H}{2} regions.  This picture is supported by
kinematically distinct emission for the two components in some
galaxies (V\'eron et al. 1997), and by aperture-dependent line ratios
on kpc scales for AGNs that may reflect different admixtures of the two emission
sources (Storchi-Bergmann 1991).  Under this scenario, we would expect
the nuclei to more strongly resemble LINERs or Seyferts in
the {\sl HST} spectra, since the small aperture should act to exclude
a part of the circumnuclear nebulosity.  

From the results reported in \S3 there is only limited support
for these predictions.  The LINER 1s and Seyfert nuclei show a modest
aperture dependence in the line ratios that appears to reflect at
least partially density effects (\S 3.2).  To the extent that
ionization gradients are present, it seems that the ionization 
often {\em decreases} on smaller scales.  If we apply to the SUNNS
measurements the same classification criteria as employed by Ho et
al. (1997a) for the Palomar observations, we find that only three
objects formally change their classification.  NGC~3992 and
NGC~4548, classified respectively as a transition object and a LINER 2
in the Palomar spectra, show higher excitation in the small aperture,
with an \ion{O}{3}/H$\beta$ ratio consistent with a borderline Seyfert
identification.  Conversely, NGC~4698, with a Palomar classification
as a Seyfert 2, shows a decrease in excitation in the SUNNS data
consistent with classification as a LINER or transition nucleus.  The
composite model for transition objects is not strongly supported by
these data.  Approximately half of the admittedly small original
sample of transition sources shows little aperture dependence in the
line ratios, and in particular these objects do not appear as pure
LINERs or Seyferts in the SUNNS measurement.  None of the narrow-line
objects reveal new evidence of broad H$\alpha$ emission when viewed at
{\sl HST} resolution (see also Gonz\'alez Delgado et al. 2004).

The behavior of the transition objects is a surprise.  The lack of
strong emission-line gradients in some sources may indicate that the
length scale over which the photoionizing energy of an accretion
source is deposited is comparable to that of a circumnuclear
star-forming zone (i.e. $\sim 100$ pc), so that the influence of both
energy sources is spatially commingled.  This would involve a
different gas distribution or geometry than is typical of the LINER 1s
and Seyferts, which tend to have highly concentrated nebulosity around
an accretion source, and may have interesting implications for
the available fuel supply.  Alternatively, a distributed ionization source,
such as turbulent mixing layers (e.g., Begelman \& Fabian 1990) or 
evolved stars (e.g. Binette et al. 1994), rather than
simply a central accretion source, may be important in these nuclei;
the magnetic field may also be energetically significant in these regions.
A distributed source is potentially supported by our finding that
transition and LINER 2 nuclei in the SUNNS sample preferentially show
evidence of a nuclear component of stars with ages $\la 1$ Gyr (Sarzi
et al. 2005; see also Gonz\'alez Delgado et al. 2004).  Investigation of
a larger sample of transition objects is important for defining the
properties of these sources and determining their nature.  The
decrease in ionization with decreasing radius in the Seyfert 2 nuclei
is likewise unexpected, and scrutiny of larger samples is needed to
understand this finding.  Integral-field spectroscopic observations could
prove useful to understanding the role of different ionizing sources at
different radii, as recently shown by Mazzuca et al. (2006) in their study
of the galaxy NGC~7742, which hosts a circumnuclear starburst and an
active nucleus of composite nature.

\subsection{Galaxy Nuclei: Ours and Others}

While the sample studied here consists of nearby emission nuclei,
there is one such source notably missing from our study, namely the
Milky Way nucleus.  Integrating the Galactic Center into a larger
context of galaxy nuclei is complicated observationally by the large
extinction to this region ($A_V \approx 31$ mag; Rieke, Rieke, \& Paul
1989), and by its close proximity, which results in an angular scale
(1\arcsec $\approx$ 0.1 pc) substantially different from extragalactic
sources.  Spectroscopic measurements obtained for both external
galaxies and the Galactic Center on the {\sl same} metric scale in the
{\sl same} bandpass remain very limited.  The SUNNS data, however,
offer a new basis for comparison by at least matching the approximate
metric scale of many measurements of the Galactic Center.

The Milky Way, like the SUNNS galaxies, is host to a low-luminosity
emission nucleus.  The rate of photoionizations by Lyman continuum
photons for the Galactic Center is tabulated as a function of radius
by Mezger, Duschl, \& Zylka (1996, $N^\prime_{Lyc}$ in their Table 6),
based on radio measurements of free-free emission.  After rescaling to
a distance to the Galactic Center of $R_0 = 8$ kpc, the results
indicate $N^\prime_{Lyc} = 2.1 \times 10^{50}$ s$^{-1}$ within $R =
0.9$ pc and $N^\prime_{Lyc} = 5.7 \times 10^{51}$ s$^{-1}$ within $R =
235$ pc.  The H$\alpha$ luminosity $L({\rm H}\alpha)$ emerging within
$R = 235$ pc predicted from recombination theory with negligible
extinction would be $\sim 8 \times 10^{39}$ ergs s$^{-1}$.  Objects
with this $L({\rm H}\alpha)$ are common in the full Palomar
spectroscopic survey, which has a slightly lower median $L({\rm
H}\alpha) = 2 \times 10^{39}$ ergs s$^{-1}$ for an equivalent radius
of $\sim 140$ pc (i.e., for a circular aperture with area equal
to the rectangular Palomar aperture; Ho et al. 1997c).  Allowing for
modest extinction of the Galactic Center emission and the somewhat
different metric apertures would bring the Milky Way value closer to
the Palomar median.  For comparison purposes with SUNNS, an estimate
of $N^\prime_{Lyc}$ within $R \approx 10$ pc for the Galaxy would be
desirable, but is not readily available; with the existing data we can
nonetheless estimate that the ratio of recombination luminosities
within $R \sim 10$ and $\sim 100$ pc is approximately one order of
magnitude, which is broadly consistent with our measured ratio for
other galaxies (Figure 2).  The Galactic Center is thus a very typical
nucleus in terms of its nebular luminosity and concentration.

The SUNNS data provide a further basis for comparing interstellar
conditions in the Galactic Center with those in other galaxy nuclei.
Nebular densities within $\sim 10$ pc tend to be rather high.  As
discussed in \S3.2, the [\ion{S}{2}] diagnostic indicates a typical
density of $\sim 10^3$ cm$^{-3}$, but other lines point to the
presence of gas with densities of $\sim 10^{4}$ cm$^{-3}$ in the
SUNNS nuclei.  Very similar results are found in the Galactic Center,
using constraints from infrared fine-structure line ratios and
emission measures (e.g., Genzel, Hollenbach, \& Townes 1994; Beckert
et al. 1996, and references therein).  The {\sl HST} spectra do not
provide quantitative constraints on the gas-phase abundances, although
our analysis of the {\sl stellar} abundances with these data indicates
that typical metallicities $Z \approx (1 - 2) \times$ Z$_\odot$ (Sarzi
et al. 2005).  This range matches estimates of the nebular metallicity
of the Galactic Center (Shields \& Ferland 1994), and estimates of
metallicity for stars in the central parsecs also suggest a typical
value of $Z \approx$ Z$_\odot$ (Ram{\'\i}rez et al. 2000; Blum et
al. 2003).

Although we cannot see the optical spectrum of the Galactic Center, we
can make some inferences about its optical spectral class from the
results noted above and other measurements.  The Galactic Center
provides compelling evidence for the presence of a black hole with a
mass $M_\bullet \approx 3 - 4 \times 10^6$ M$_\odot$ (e.g., Sch\"odel
et al. 2003; Ghez 2004); this mass and the stellar velocity dispersion
$\sigma_\ast$ for the Milk Way bulge are in reasonable accord with the
$M_\bullet - \sigma_\ast$ correlation defined by other galaxies.  The
central black hole is apparently accreting, based on its appearance as
the compact radio source Sgr A$^\ast$ (Balick \& Brown 1974) and an
associated X-ray source (Baganoff et al. 2003).  The accretion
luminosity is, however, very feeble ($\la 10^{37}$ ergs s$^{-1}$;
Narayan et al. 1998, and references therein), and thus insufficient to
power a significant fraction of the nebular emission.  The nebular
emission on scales from $\sim 1$ pc (e.g., Lutz et al. 1996) to $\sim
60$ pc (e.g., Simpson et al. 1999) is characterized by low-ionization
lines.  The lack of both a strong compact X-ray source and
high-ionization nebulosity implies that the Milky Way is not a Seyfert
nucleus.  Likewise, the central parsec of the Galaxy does not exhibit
strong, quasar-like permitted lines from high-velocity gas; the
Galactic Center is not a broad-line (Type 1) nucleus.

Would our counterparts in other galaxies regard the Milky Way as any
sort of AGN, of the types found in the Palomar spectroscopic survey?
If so, our Galaxy would have to be a LINER or \ion{H}{2}/LINER
transition object.  A LINER generated via photoionization is excluded,
based on the absence of a significant compact hard X-ray source to
power the nebulosity, as well as from detailed aspects of the infrared
and radio emission lines (Shields \& Ferland 1994; Lutz et al. 1996).
Recent observations with {\sl Chandra} reveal diffuse hard X-ray
emission from the central $R \approx 60$ pc, with a luminosity of
$\sim 2 \times 10^{36}$ ergs s$^{-1}$ (Muno et al. 2004).  While
insufficient to power significant nebular emission directly, the hard
X-rays are of uncertain origin, and would require a kinetic energy
input of $\sim 10^{40}$ ergs s$^{-1}$ if generated in (unbound)
thermal plasma (see Muno et al. 2004 for details).  If a fraction of
this mechanical energy were transferred to the nebular plasma, it
could modify the excitation of the gas, resulting in enhancement of
optical collisionally excited lines.  However, the nebular gas is
clearly not heavily shocked; the plasma is not highly ionized, and at
least within $\sim 6$ pc of the center, the electron temperature as
revealed by radio line-to-continuum ratios is only $7000 \pm 500$ K
(Roberts \& Goss 1993).  While this value may be slightly higher than
expected from simple stellar photoionization (Shields \& Ferland
1994), it appears that the gas is not heated sufficiently to power
optical forbidden lines at the level seen in LINERs.

We conclude that the center of the Milky Way would appear
spectroscopically as an \ion{H}{2} nucleus or possibly a transition
object for an extragalactic optical observer.  As in most of the SUNNS
galaxies (Sarzi et al. 2005), the inner stellar bulge of our Galaxy is
predominantly old (Blum et al. 2003; van Loon et al. 2003; although
cf. Figer et al. 2004), but some star formation has occurred in recent
episodes, and in the case of the Milky Way, the ultraviolet radiation
from normal massive stars apparently dominates the nebular ionization.
For galaxies like our own, with a Hubble type of approximately Sbc,
the Palomar spectroscopic survey found that the fractions hosting
\ion{H}{2} and transition nuclei are $\sim 50$\% and $\sim 20$\%,
respectively (Ho et al. 1997c).  In its nebular excitation, star
formation history, and relation to the larger Galaxy, the Milky Way
again has a very typical nucleus.

\section{Summary and Conclusions}

Ground-based spectroscopic surveys indicate that the majority of
galaxies host emission-line nuclei, and the SUNNS results provide new
insights into the structure and energetics of these nebular sources.
The findings presented here indicate that the emitting gas is
spatially distributed on scales of $\sim 100$ pc, with
accretion-powered objects showing a significant degree of
concentration in the observed emission.  Within these regions, modest
radial gradients in optical emission-line ratios are seen; density
gradients evidently play an important role in driving these trends,
although there is some evidence as well for ionization gradients in
the plasma.  LINER/\ion{H}{2} transition nuclei show only mixed
indications of a simple composite spatial structure, contrary to
expectations, suggesting that distributed sources of ionization beyond
normal O-type stars may be involved in powering these common objects.

These findings provide a framework for gauging the extent to which the
center of the Milky Way is a typical galaxy nucleus.  The Galactic
Center is very similar to other emission-line nuclei in terms of its
nebular luminosity, surface brightness distribution, and ionization.
Combining the present results with other published findings, it is
additionally clear that the Galactic Center resembles the nuclei of
other systems in terms of its metallicity and stellar population, and
likewise its black hole mass and probable spectroscopic class,
considered in relation to the bulge properties of the host galaxy.  As
a laboratory for the study of galaxy nuclei, the center of the Milky
Way thus provides a very representative example.

\acknowledgements
We thank the referee for constructive comments.
Support for this research was provided by NASA through grant GO-7361
from the Space Telescope Science Institute, which is operated by the
Association of Universities for Research in Astronomy, Inc., under
NASA contract NAS5-26555.

\begin{deluxetable}{cccc}
\tabletypesize{\scriptsize}
\tablewidth{0pt}
\tablenum{1}
\tablecaption{Galaxy Sample}
\tablehead{
Name & Distance\tablenotemark{a} & Hubble Type\tablenotemark{b} 
& Ground-based \\
     & (Mpc)    & & spectral class\tablenotemark{c} 
}
\startdata
NGC  278 & 11.8 & SAB(rs)b  & H     \\ 
NGC 2787 & 13.0	& SB(r)0+   & L1.9  \\ 
NGC 3351 &  8.1	& SB(r)b    & H     \\ 
NGC 3368 &  8.1	& SAB(rs)ab & L2    \\ 
NGC 3489 &  6.4	& SAB(rs)0+ & T2/S2 \\ 
NGC 3982 & 17.0	& SAB(r)b:  & S1.9  \\ 
NGC 3992 & 17.0	& SB(rs)bc  & T2:   \\ 
NGC 4138\tablenotemark{d} & 17.0 & SA(r)0+ & S1.9  \\ 
NGC 4143 & 17.0	& SAB(s)0   & L1.9  \\ 
NGC 4203 &  9.7	& SAB0-:    & L1.9  \\ 
NGC 4245 &  9.7	& SB(r)0/a: & H     \\ 
NGC 4314 &  9.7	& SB(rs)a   & L2    \\ 
NGC 4321 & 16.8	& SAB(s)bc  & T2    \\ 
NGC 4380 & 16.8	& SA(rs)b?  & H     \\ 
NGC 4435 & 16.8	& SB(s)0    & T2/H: \\ 
NGC 4450 & 16.8	& SA(s)ab   & L1.9  \\ 
NGC 4459 & 16.8	& SA(r)0+   & T2:   \\ 
NGC 4477 & 16.8	& SB(s)0?   & S2    \\ 
NGC 4501 & 16.8	& SA(rs)b   & S2    \\ 
NGC 4548 & 16.8	& SB(rs)b   & L2    \\ 
NGC 4596 & 16.8	& SB(r)0+   & L2::  \\ 
NGC 4698 & 16.8	& SA(s)ab   & S2    \\ 
NGC 4800 & 15.2	& SA(rs)b   & H     \\ 
NGC 5055 &  7.2	& SA(rs)bc  & T2    \\ 
\enddata
\tablenotetext{a}{From Tully (1988).}
\tablenotetext{b}{From de Vaucouleurs et al. (1991)}
\tablenotetext{c}{From Ho et al. (1997a).  For sources with two
possible classifications, the first listed class is marginally preferred,
and adopted here when objects are sorted by type. Colons and double-colons
indicate uncertain and highly uncertain classifications, respectively.}
\tablenotetext{d}{Due to a pointing error during the observations of NGC~4138,
this galaxy was dropped from the analysis.}
\end{deluxetable}

\begin{deluxetable}{ccccccccccc}
\tabletypesize{\scriptsize}
\rotate
\tablewidth{0pt}
\tablenum{2}
\tablecaption{Narrow Emission Line Measurements}
\tablehead{
Galaxy & [O II]        &            &               & [O III] &
         [O I]         &             & [N II]       & [S II]  &
         [S II]        & EW(H$\alpha$) \\

Name   & $\lambda$3727 & H$\gamma$  & H$\beta$      & $\lambda$5007 &
         $\lambda$6300 & H$\alpha$  & $\lambda$6583 & $\lambda$6716 &
         $\lambda$6731 & (\AA )     \\

(1)    & (2)           & (3)        & (4)           & (5)           &
         (6)           & (7)        & (8)           & (9)           &
         (10)          & (11)  \\
}

\startdata
NGC  278 &  $<23$     &  $<23$     &  $<23$     &  $<23$      
         &  $<25$     &  $<25$     &  $<25$     &  $<25$     &  $<25$     & $<1.0$ \\   
NGC 2787\tablenotemark{a} & 390$\pm$40 &  106$\pm$11  & 200$\pm$20 & 280$\pm$30 
         & 118$\pm$12 & 180$\pm$20 & 500$\pm$50 & 200$\pm$20 & 210$\pm$20 & 6.3$\pm$0.9 \\
NGC 3351 &  $<9$      &  $<9$      & 22$\pm$3   &  25$\pm$3
         &  $<54$     & 126$\pm$13 & 71$\pm$7   &  $<18$     & $<18$      & 19$\pm$3 \\
NGC 3368 & 67$\pm$11  &  $<32$     &  $<32$     &  $<32$
         &  $<34$     & 117$\pm$12 & 139$\pm$14 & 59$\pm$11  & 59$\pm$11  & 1.7$\pm$0.2\\
NGC 3489 &  $<39$     &  $<39$     &  $<39$     &   $<39$     
         &  $<57$     &  $<57$     &  $<57$     &   $<57$    &  $<57$     & $<0.6$ \\
NGC 3982 & 560$\pm$60 & 130$\pm$15 & 240$\pm$40 & 5000$\pm$500 
         & 620$\pm$60 & 1400$\pm$140 & 1380$\pm$140 & 310$\pm$30 & 460$\pm$50 & 180$\pm$20 \\
NGC 3992 & 61$\pm$6   & 12$\pm$4   & $<11$      & 36$\pm$4
         &  $<18$     & 44$\pm$6   & 87$\pm$9   & 37$\pm$6   & 32$\pm$6   & 3.0$\pm$0.4 \\
NGC 4143\tablenotemark{a} & 690$\pm$70 & $<170$ & 400$\pm$40 & 900$\pm$90
         & 750$\pm$70 & 390$\pm$40 & 1070$\pm$110 & 340$\pm$30  & 340$\pm$30 & 11$\pm$2 \\
NGC 4203\tablenotemark{a} & 950$\pm$100 & 450$\pm$40 & 900$\pm$90   & 1700$\pm$170
         & 1070$\pm$110 & 450$\pm$40 & 1080$\pm$110 & 250$\pm$30  & 340$\pm$30 & 10$\pm$1 \\
NGC 4245 &  $<15$     &  $<15$     &  $<15$     &  $<15$      
         &  $<28$     &  $<28$     &  $<28$     &  $<28$     &  $<28$     & $<1.2$ \\
NGC 4314 &  $<13$     &  $<13$     &  $<13$     &  $<13$     
         &  $<22$     & 32$\pm$7   & $<22$      &  $<22$     &  $<22$     & 2.5$\pm$0.6 \\
NGC 4321 & 100$\pm$10 & 53$\pm$6   & 140$\pm$14 & 47$\pm$6   
         & 106$\pm$13 & 740$\pm$70 & 700$\pm$70 & 112$\pm$11 & 132$\pm$13 & 37$\pm$5 \\
NGC 4380 &  $<9$      &  $<9$      &  $<9$      &  $<9$     
         &  $<15$     &  $<15$     &  $<15$     &  $<15$     &  $<15$     & $<2.5$ \\
NGC 4435 &  $<14$     &  $<14$     &  $<14$     &  $<14$     
         &  $<22$     & 28$\pm$7   & 63$\pm$7   &  $<22$     &  $<22$     & 1.9$\pm$0.4 \\
NGC 4450\tablenotemark{a} & 800$\pm$80 & 220$\pm$20 & 400$\pm$40 & 800$\pm$80
         & 500$\pm$50 & 400$\pm$40 & 770$\pm$80 & 420$\pm$40 & 460$\pm$50 & 12$\pm$2 \\
NGC 4459 & 42$\pm$7   &  $<22$     &  $<22$     & 27$\pm$7     
         &  $<36$     & 63$\pm$12  & 250$\pm$20 &  $<36$     & 38$\pm$12  & 1.4$\pm$0.2 \\
NGC 4477 & 86$\pm$9   & 14$\pm$4   & 28$\pm$4   & 82$\pm$8
         & 26$\pm$9   & 110$\pm$11 & 240$\pm$20 & 62$\pm$9   & 62$\pm$9   & 5.6$\pm$0.8 \\
NGC 4501 & 98$\pm$10  &  $<14$     & 36$\pm$5   & 163$\pm$16 
         & 54$\pm$9   & 152$\pm$15 & 360$\pm$40 & 96$\pm$10  & 92$\pm$9   & 5.7$\pm$1.1 \\
NGC 4548 & 55$\pm$6   &  $<10$     & $<10$      & 37$\pm$4
         & 23$\pm$7   & 48$\pm$7   & 110$\pm$11 & 48$\pm$7   & 40$\pm$7   & 3.2$\pm$0.5 \\
NGC 4596 & $<15$      & $<15$      & $<15$      & $<15$      
         & $<23$      & 25$\pm$8   & 69$\pm$8   & $<23$      & $<23$      & $<0.7$ \\
NGC 4698 & 90$\pm$9   & 40$\pm$5   & 60$\pm$6   & 149$\pm$15
         & $<39$      & 150$\pm$15 & 200$\pm$20 & 60$\pm$6   & 70$\pm$7   & 8.9$\pm$1.3 \\
NGC 4800 & $<13$      & $<13$      & $<13$      & $<13$      
         & $<20$      & $<20$      & $<20$      & $<20$      & $<20$      & $<1.0$   \\
NGC 5055 & $<76$      & $<76$      & $<76$      & $<76$     
         & $<71$      & $<71$      & $<71$      & $<71$      & $<71$      & $<0.6$ \\
\enddata
\tablenotetext{a} {See \S 3.3 for discussion of systematic errors in H$\alpha$
fluxes for these sources.}
\tablecomments{
Col. (1): Galaxy name. Cols. (2)--(10): Narrow-line fluxes in units of 
$10^{-17}$ ergs s$^{-1}$ cm$^{-2}$; upper limits are 3$\sigma$. 
Col. (11): Equivalent width of narrow H$\alpha$. \\
}
\end{deluxetable}

\begin{deluxetable}{ccccc}
\tablewidth{0pt}
\tablenum{3}
\tablecaption{Broad H$\alpha$ Properties\tablenotemark{a}}
\tablehead{
Galaxy & \multicolumn{2}{c}{Palomar Measurement\tablenotemark{b}} & \multicolumn{2}{c}
{SUNNS Measurement\tablenotemark{c}}\\
       & $\log F({\rm H}\alpha)$ & $f_{blend}$ & $\log F({\rm H}\alpha)$ & $f_{blend}$ 
}
\startdata
NGC 2787 & $-13.55$ & 0.35 & $\le -14.02$ & $\le 0.53$ \\
NGC 3982 & $-13.85$ & 0.12 & $\le -15.05$ & $\le 0.03$ \\
NGC 4143 & $-13.41$ & 0.45 & $\le -13.66$ & $\le 0.55$ \\
NGC 4203\tablenotemark{d} & $-13.46$ & 0.34 & $\le -13.54$ ($-12.83$) & $\le 0.61$ (0.89) \\
NGC 4450\tablenotemark{d} & ... & 0.20 & $\le -14.00$ ($-13.20$) & $\le 0.41$ (0.82) \\
NGC 4501 & $-14.02$ & 0.09 & $    -14.66$ & $ 0.26$ \\

\enddata
\tablenotetext{a}{Fluxes are in units of ergs s$^{-1}$ cm$^{-2}$ and are not
corrected for extinction.  The quantity $f_{blend}$ is the fraction of the total
H$\alpha$+[\ion{N}{2}] flux contributed by the broad H$\alpha$ feature.}
\tablenotetext{b}{From Ho et al. 1997b.}
\tablenotetext{c}{Upper limits reflect possible systematic uncertainties
in profile decomposition (\S 3.3).  Uncertainty in the SUNNS measurement
for NGC~4501 is $\sim 100$\% .}
\tablenotetext{d}{Fluxes for ``normal'' Gaussian component as reported by 
Ho et al.  1997a.  Additional, very broad components were detected in
the SUNNS spectra, and results with this component included are listed
in parentheses; see Shields et al.  (2000, NGC~4203) and Ho et al. (2000,
NGC~4450).}
\end{deluxetable}

\end{document}